\documentclass[12pt]{iopart}
\usepackage{graphicx}
\usepackage{bm}
\usepackage{amssymb}
\usepackage{color}
\usepackage{latexsym}

\begin{document}
\title[Interaction Between Motor Domains in Kinesins]{Interaction Between Motor Domains Can Explain the Complex Dynamics of Heterodimeric Kinesins}
\author{Rahul Kumar Das, Anatoly B. Kolomeisky}
\address{Department of Chemistry, Rice University, Houston, Texas 77005}

\begin{abstract}
Motor proteins are active enzyme molecules that play a crucial role in many biological processes. They transform the chemical energy into the mechanical work and move unidirectionally along rigid cytoskeleton filaments. Single-molecule experiments suggest that motor proteins, consisting of two  motor domains, move in a hand-over-hand mechanism when each subunit changes between trailing and leading positions in alternating steps,  and these subunits do not interact with each other. However, recent experiments on heterodimeric kinesins suggest that the motion of motor domains is not independent, but rather strongly coupled and coordinated, although the mechanism of these interactions are not known. We propose a simple discrete stochastic model to describe  the dynamics of homodimeric and heterodimeric two-headed  motor proteins. It is argued that interactions between motor domains modify free energy landscapes of each motor subunit, and motor proteins still move via the hand-over-hand mechanism but with different transitions rates. Our calculations of biophysical properties agree with experimental observations. Several ways to test the theoretical model are proposed.
 
\end{abstract}

\ead{tolya@rice.edu}

\maketitle

\section{Introduction}

Several classes of enzyme molecules that convert chemical energy into mechanical motion are called motor proteins, or molecular motors. In recent years, these proteins have attracted a significant attention  because of their importance for multiple biological processes \cite{lodish,bray,howard,schliwa,AR}. Motor proteins, such as kinesins, myosins, dyneins, polymerases and helicases, move in a linear fashion along the rigid biopolymers (actin filaments, microtubules, DNA and RNA molecules). Typically, a fuel for the motion of these nanomotors  comes from the hydrolysis of adenosine triphosphate (ATP) or related compounds. Although some progress in understanding the mechanisms of motor proteins has been achieved \cite{howard,schliwa,AR}, there are still many unresolved issues. One of the most important fundamental questions concerning motor proteins  is how different domains of these enzymes coordinate and regulate their complex dynamics and functions. The goal of this paper is to address some aspects of this issue from the theoretical point of view.

The enzymatic activity of motor proteins takes place in the so-called motor subunits that contain ATP-binding sites. Motor proteins typically have several such domains. The functioning of molecular motors strongly depends on the relative position and dynamics of these subunits \cite{howard, AR}. Two possible mechanisms for two-headed motor proteins have been proposed: an inchworm motion and a hand-over-hand mechanism \cite{howard,kolomeisky05}. In the inchworm mechanism one motor domain is always in the leading position, while another one always trails. However, in the hand-over-hand mechanism the motor domains alternate their leading and trailing positions as the motor protein molecule proceeds along the filament track. Single-molecule experiments that utilized fluorescent imaging with one-nanometer accuracy (FIONA) have shown that individual double-headed kinesins, myosins V and VI, and cytoplasmic dyneins step in the hand-over-hand fashion \cite{yildiz03,snyder04,yildiz04,toba06,reck-peterson06}. Thus, this mechanism explains stepping dynamics of the majority of motor protein species. 

In the current version of the hand-over-hand mechanism  it is assumed that two heads move independently from each other, i.e., when the trailing motor subunit moves its dynamics is not affected by the presence of another motor subunit. Then a mean dwell time to advance one step forward for a heterodimeric motor protein with two heads labeled as $A$ and $B$ is given by
\begin{equation}
\tau_{A/B}=\frac{1}{2} (\tau_{A/A} + \tau_{B/B}),
\end{equation}
where $\tau_{A/A}$ and $\tau_{B/B}$ are mean  dwell times for homodimeric $A/A$ and $B/B$ motor proteins, respectively. The corresponding relation for the velocity can be written as
\begin{equation}
V_{A/B}=\frac{2 V_{A/A} V_{B/B}} {V_{A/A} + V_{B/B}}.
\end{equation}
However recent single-molecule investigations of dynamics of kinesins \cite{kaseda02} do not support these relations, and, consequently, the independence of two motor domains during the motion is put in a doubt.  In these experiments force-velocity curves and enzymatic activities have been measured for different homodimeric and heterodimeric kinesins. Surprisingly, it was shown that the velocity of the heterodimeric kinesin with a mutation in one of the motor heads is not given by Eq. (2). It was suggested that the two heads strongly influence the dynamic and enzymatic properties of each other, although the  mechanism was not specified. In this paper we present a simple discrete stochastic model that might explain several aspects of the complex dynamics of heterodimeric kinesins. Our main idea is that motor domains interact with each other and significantly modify the overall dynamics.  

Theoretical investigations of molecular motors follow several approaches that include continuum ratchets \cite{julicher97,bustamante01}, discrete stochastic models \cite{AR} and computer simulations \cite{li04,hyeon07}. In this work we utilize a discrete stochastic approach because it conveniently  provides explicit expressions for dynamic properties, and it is able to describe successfully different aspects and trends of motor proteins transport \cite{AR,kolomeisky05,FK,kolomeisky01,KF00a,KF00b,FK2,KF03,stukalin05,fisher05}.

\section{Theoretical Model}

In our theoretical model it is assumed that a  kinesin protein molecule steps from a given binding site to the next one at a distance $d=8.2$ nm along the microtubule through $N$ intermediate biochemical states. Different kinetic schemes for the motion of homodimeric and heterodimeric kinesin molecules are shown in Fig. 1. Similar  approach has been used successfully to describe  dynamic properties of kinesins and myosins V \cite{AR,FK,FK2,KF03}.

\begin{figure}[tbp]
\centering
\includegraphics[scale=0.5,clip=true]{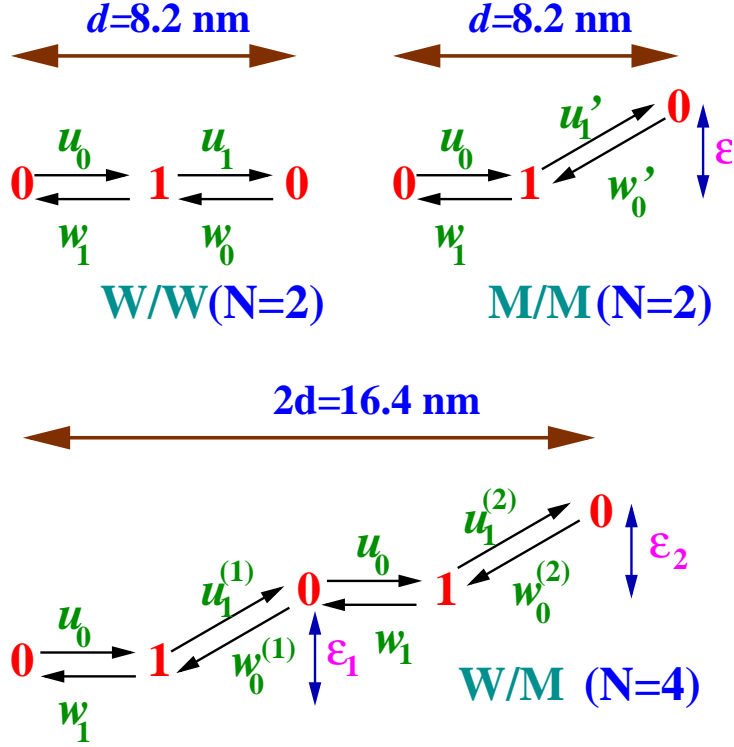}
\caption{General schematic view of discrete stochastic models for kinesins. W/W labels the homodimeric motor protein with both wild-type motor heads, M/M corresponds to the homodimeric kinesins with mutations in both motor domains, while W/M describes the heterodimeric motor proteins with mutation in only of the heads. Parameters $\varepsilon$, $\varepsilon_{1}$ and $\varepsilon_{2}$ describe free energy changes due to mutations relative to the wild-type motor proteins.}
\label{fig1}
\end{figure}

Let us label homodimeric motor proteins of wild type as W/W, homodimeric motor proteins with mutations in both heads as M/M, and heterodimeric proteins with only one mutated motor domain as W/M. Although the motion of motor proteins includes many complex biochemical and biophysical processes, in the simplest approximation it is reasonable to use only  two-state ($N=2$) discrete stochastic models to describe the dynamics of both homodimeric kinesins: see Fig. 1. The motor protein molecule can jump forward from a state 0 to a state 1 with the rate $u_{0}$, and this transition corresponds to ATP binding, yielding
\begin{equation}
u_{0}=k_{0}\mbox{[ATP]},
\end {equation}
where $k_{0}$ is a rate constant. The reverse transition is given by the rate $w_{1}$. The forward and backward transitions between the state 1 and the state 0 on the next binding site, with the rates $u_{1}$ and $w_{0}$ respectively, describe several biochemical processes, such as  ATP hydrolysis and release of hydrolysis products, but combine them in one step. We assume that mutations do not affect strongly  ATP binding process, but only the enzymatic functions are changed, which leads to a different pair of transition rates ($u_{1}'$ and $w_{0}'$) for the mutated molecule (see Fig. 1). Experiments show that some mutations decrease the binding affinity of motor proteins to microtubules, thus leading to decrease in enzymatic activity \cite{kaseda02}.  The transition rates for W/W and M/M motor proteins are related via the detailed balance condition,
\begin{equation}\label{eq.rates1}
\frac{u_{1}'}{w_{0}'}=\frac{u_{1}}{w_{0}} \exp(-\varepsilon/k_{B}T).
\end{equation}
The parameter $\varepsilon$  describes the effect of mutation on the enzymatic properties of the motor protein, i.e., how the ATP hydrolysis for the mutated homodimer is thermodynamically less favorable in comparison with that of wild-type homodimer. Microscopic origin of this parameter is the result of complex interactions between two motor domains and between  motor heads and the microtubule track.

The situation is more complex for heterodimeric W/M kinesins, as illustrated in Fig. 1. Two motor domains  and interactions between them are different from the cases of homodimer kinesins. As a result we have two different sets of rates to model ATP hydrolysis by the wild head ($u_{1}^{(1)}$ and $w_{0}^{(1)}$) and by the mutated head ($u_{1}^{(2)}$ and $w_{0}^{(2)}$). These rates are also related via detailed balance conditions,
\begin{equation}\label{eq.rates2}
\frac{u_{1}^{(1)}}{w_{0}^{(1)}}=\frac{u_{1}}{w_{0}} \exp(-\varepsilon_{1}/k_{B}T), \quad \frac{u_{1}^{(2)}}{w_{0}^{(2)}}=\frac{u_{1}}{w_{0}} \exp(-\varepsilon_{2}/k_{B}T).
\end{equation}
It is important to note that generally, $\varepsilon_{1} \neq \varepsilon_{2} \neq \varepsilon$, because of different interactions between the motor domains. Then, assuming that the hand-over-hand mechanism still is a valid description for the stepping of individual molecules, the dynamics of W/M kinesin molecules can be described by $N=4$-state model with step size equal to 2$d$=16.4 nm. 

The explicit expressions for the rates can be obtained from Eqs. (\ref{eq.rates1}) and (\ref{eq.rates2}):
\begin{equation}
u_{1}'= u_1\gamma^{-\alpha}, u_{1}^{(1)}= u_1 \gamma_{1}^{-\alpha_{1}}, u_{1}^{(2)}= u_1\gamma_{2}^{-\alpha_{2}};
\end {equation}
and
\begin{equation}
w_{0}'= w_0\gamma^{1-\alpha}, w_{0}^{(1)}= w_0 \gamma_{1}^{1-\alpha_{1}}, w_{0}^{(2)}= w_0 \gamma_{2}^{1-\alpha_{2}},
\end {equation}
where we defined
\begin{equation}
\gamma= \exp\left({\frac{\epsilon}{kT}}\right), \gamma_{1}= \exp\left({\frac{\varepsilon_{1}}{kT}}\right), \gamma_{2}= \exp\left({\frac{\varepsilon_{2}}{kT}}\right).
\end {equation}
Parameters $\alpha$, $\alpha_{1}$  and $\alpha_{2}$ are energy-distribution factors that describe how the free energies change due to the mutation affects corresponding to the forward and the backward transitions. For simplicity, we assume that $\alpha=\alpha_{1}=\alpha_{2}$, although more general situations can be easily analyzed.

The advantage of using discrete stochastic models is the fact that all stationary-state dynamic properties of motor proteins, such as mean velocities, mean dispersions and stall forces, can be obtained exactly for {\it any} number of intermediate states $N$ in terms of the forward ($u_{j}$) and backward ($w_{j}$) transition rates \cite{AR,KF00a,FK2}. Specifically, exact expressions for the mean velocity can be presented in the following form,
\begin {equation}\label{eq.vel}
V=\frac{d}{R_{N}}\left[1-{\prod_{j=0}^{N-1}{\frac{w_j}{u_j}}}\right]
\end {equation}
 where $d$ is the step size (equal to 8.2 nm for $N=2$ model and 16.4 nm for $N=4$ model), and the auxiliary functions $R_N$ are given by 
\begin {equation} 
R_{N}=\sum_{j=0}^{N-1} r_{j}, \quad r_j=\frac{1}{u_j}\left[1+\sum_{k=1}^{N-1} \prod_{i=j+1}^{j+k} \frac{w_{i}}{u_{i}}\right].
\end {equation}
Specifically, for the velocity of homodimeric motor proteins we obtain
\begin{equation}
V(W/W)=d \left[\frac{u_{0}u_{1}-w_{0}w_{1}}{u_{0}+u_{1}+w_{0}+w_{1}}\right]
\end {equation}
for W/W kinesins, and 
\begin{equation}
V(M/M)=d \left[\frac{u_{0}u_{1} \gamma^{-\alpha}-w_{0}w_{1}\gamma^{1-\alpha}}{u_{0}+u_{1}\gamma^{-\alpha} +w_{0}\gamma^{1-\alpha}+w_{1}}\right]
\end {equation}
for M/M kinesins. For heterodimeric motor proteins Eq. (\ref{eq.vel}) yields
\begin {equation}
V(W/M)=2d \left[\frac{u_{0}^{2}u_{1}^{2} (\gamma_{1} \gamma_{2})^{-\alpha} -w_{0}^{2}w_{1}^{2}(\gamma_{1} \gamma_{2})^{1-\alpha}}{A}\right],
\end {equation}
with the parameter $A$ given by
\begin {eqnarray}
A &=&(\gamma_{1}^{-\alpha}+\gamma_{2}^{-\alpha})u_{0}u_{1}(u_{0}+w_{1})+(\gamma_{1}^{1-\alpha}+\gamma_{2}^{1-\alpha})w_{0}w_{1}(u_{0}+w_{1})+ \\ \nonumber 
 & & (\gamma_{1} \gamma_{2})^{-\alpha}(\gamma_{1} + \gamma_{2})u_{1}w_{0}(u_{0}+w_{1})+2 (\gamma_{1} \gamma_{2})^{-\alpha}(u_{0}u_{1}^{2}+w_{0}^{2}w_{1}\gamma_{1} \gamma_{2}).
\end {eqnarray}

The general expression for dispersion in sequential discrete stochastic models can be written in the following form \cite{AR,KF00a,FK2},
\begin {equation}
D=(d/N) \left\{ [V S_{N} + d U_{N}]/(R_{N})^{2}-(N+2)V/2 \right\},
\end {equation}
where
\begin {equation} 
S_{N}=\sum_{j=0}^{N-1} s_{j} \sum_{k=0}^{N-1} r_{k+j+1}, \quad U_{N}=\sum_{j=0}^{N-1} u_{j}r_{j}s_{j}, \quad s_{j}=\frac{1}{u_j}\left[1+ \sum_{k=1}^{N-1}{\prod_{i=j-1}^{j-k}{\frac{w_{i+1}}{u_{i}}}}\right].
\end {equation}
The explicit equations for dispersions of W/W, M/M and W/M kinesins can be obtained in the way similar to the velocities, however, these expressions are quite bulky and they will not be presented here.

When the  motor protein is a subject of external loads,  the resisting force that completely stops the molecule is called a stall force, $F_{S}$. For general $N$-state sequential discrete-stochastic models the stall force can be written as \cite{AR,FK}   
\begin{equation}
F_{S}=\frac{k_{B}T}{d}\ln \left( \prod_{j=0}^{N-1} \frac{w_{j}}{u_{j}} \right).                        
\end {equation}
For homodimeric kinesins  our model predicts the following stall forces,
\begin{equation}
F_{S}(W/W)=\frac{k_{B}T}{d}\ln \frac{u_{0}u_{1}}{w_{0}w_{1}}, \quad F_{S}(M/M)=\frac{k_{B}T}{d}\ln \frac{u_{0}u_{1}}{w_{0}w_{1}\gamma}.
\end{equation} 
Comparing these equations, we obtain
\begin{equation}\label{eq.stall1}
F_{S}(M/M)=F_{S}(W/W)-\varepsilon/d.
\end{equation} 
For heterodimeric kinesins the stall force is given by
\begin{equation}
F_{S}(W/M)=\frac{k_{B}T}{2d}\ln \frac{u_{0}^{2}u_{1}^{2}}{w_{0}^{2}w_{1}^{2}\gamma_{1}\gamma_{2}},
\end{equation}
which leads to 
\begin{equation}\label{eq.stall2}
F_{S}(W/M)=F_{S}(W/W)-(\varepsilon_{1}+\varepsilon_{2})/2d.
\end{equation} 
Eqs. (\ref{eq.stall1}) and (\ref{eq.stall2}) provide a simple physical interpretation and a method of estimating  parameters $\varepsilon$, $\varepsilon_{1}$ and $\varepsilon_{2}$.

External force $F$ also strongly modifies transitions rates \cite{AR,FK}
\begin{equation}
u_{j}(F)=u_j(0)\exp{\left(-\frac{\theta_{j}^{+}Fd}{k_{B}T}\right)},
\end {equation}
\begin{equation}
w_{j}(F)=w_j(0)\exp{\left(+\frac{\theta_{j}^{-}Fd}{k_{B}T}\right)},
\end {equation}
where $\theta_{j}^{\pm}$ are load-distribution factors that describe how the external load changes the energy activation barriers for the forward and backward biochemical transitions from the state $j$. Load-distribution factors are related via 
\begin{equation}
{\sum_{j=0}^{N-1}}(\theta_{j}^{+}+\theta_{j}^{-})=1.
\end {equation}

\section {Results and Discussion}

In the experimental work of Kaseda et al. \cite{kaseda02} the coordination of two heads for different homodimeric and heterodimeric kinesin molecules has been investigated using microtubule-gliding assays and optical trapping spectroscopy. Different homodimeric and heterodimeric motor proteins were prepared by  mutations in the motor domains that affect microtubule-binding region. It was found that dynamic properties of heterodimeric proteins with one mutated head could not be described by independent hand-over-hand stepping mechanism. 

To analyze experimental data we consider kinesins with only one type of mutation, although our method can be easily applied to different molecular motor species. Specifically, it was shown \cite{kaseda02} that at [ATP]=1 mM a wild-type homodimer travels with the stationary  velocity  $V(W/W)=679 \pm 59$ nm/s, and it produces the maximum stall force of $F_{S}(W/W)=6.3 \pm 0.9$ pN. When the mutation labeled L12 affects both motor heads the resulting homodimeric M/M kinesin does not attach to microtubules, indicating zero velocity and stall force. However, surprisingly, heterodimer with the same mutation L12 in one of the motor heads can move with the velocity $V(W/M)=101 \pm 25$ nm/s, while exerting the maximal stall force of $F_{S}(W/M)=0.8 \pm 0.2$ pN. Then, from Eqs. (\ref{eq.stall1}) and (\ref{eq.stall2}) we obtain
\begin{equation}
\varepsilon=12.6 \pm 1.8 \mbox{ k}_{\mbox{B}}\mbox{T}, \quad \varepsilon_{1}+\varepsilon_{2}=11 \pm 4.4 \mbox{ k}_{\mbox{B}}\mbox{T}.
\end {equation}
The important result is that $ \varepsilon_{1} + \varepsilon_{2} < 2 \varepsilon $, which indicates that biochemical properties of mutated and wild-type motor heads in the W/M kinesin  differ from the corresponding properties in the W/W and M/M motor proteins, supporting our idea of modifying free-energy landscapes for motor proteins via interaction between motor subunits.

After systematically exploring parameter space, we found that all experimental data for kinesins with L12 mutation can be well described by the following parameters,
\begin{eqnarray}
k_{0}=1.2 \mu\mbox{M}^{-1} \mbox{s}^{-1}, \  u_{1}=90\mbox{s}^{-1}, \  w_{1}=10\mbox{s}^{-1}, \  w_{0}= 0.05\mbox{s}^{-1}, \nonumber \\
\alpha=0.14, \epsilon = 12.3\mbox{ k}_{\mbox{B}}\mbox{T},\ \varepsilon_{1}=\varepsilon_{2}=10.6 \mbox{ k}_{\mbox{B}}\mbox{T}.
\end{eqnarray}
Note that obtained transition rates are similar to parameters previously utilized to describe the dynamics of single kinesins, and they are consistent with chemical kinetic experimental results \cite{FK2}. In addition, in our calculations we also used load-distribution factors reported earlier  \cite{FK2},
\begin{eqnarray}
\theta_{0}^{+}=0.135, \  \theta_{1}^{-}=0.080,\   \theta_{1}^{+}=0.035, \  \theta_{0}^{-}=0.750.
\end{eqnarray}

The results of theoretical calculations for the velocities of homodimeric and heterodimeric kinesins at different conditions are presented in Fig. 2. The effect of external load on the motor protein dynamics is shown in Fig. 2a. The force slows down the motion of all  motor proteins, as expected. It is found that the stall forces $F_{S}$ for W/W and W/M kinesins are equal to 6.2 pN and 0.8 pN respectively, which are in excellent agreement with experimentally measured values \cite{kaseda02}. The force-velocity curve for heterodimer is essentially linear, while for homodimeric W/W kinesins it deviates from the linear dependence. Again, this theoretical result agrees well with experimental observations (compare with Fig. 4b from Ref. \cite{kaseda02}). The dependence of the velocity on the concentration of ATP (see Fig. 2b) shows typical Michaelis-Menten character, i.e., a linear behavior at small concentrations of ATP and a saturation at large [ATP]. One of the important theoretical predictions of our model is that the  heterodimeric kinesin with L12 mutation in one of the motor heads will not function ($V=0$) even without external resisting force ($F=0$) for ATP concentrations smaller than 0.186 mM. This theoretical prediction can be easily  tested in experiments.

\begin{figure}[tbp]
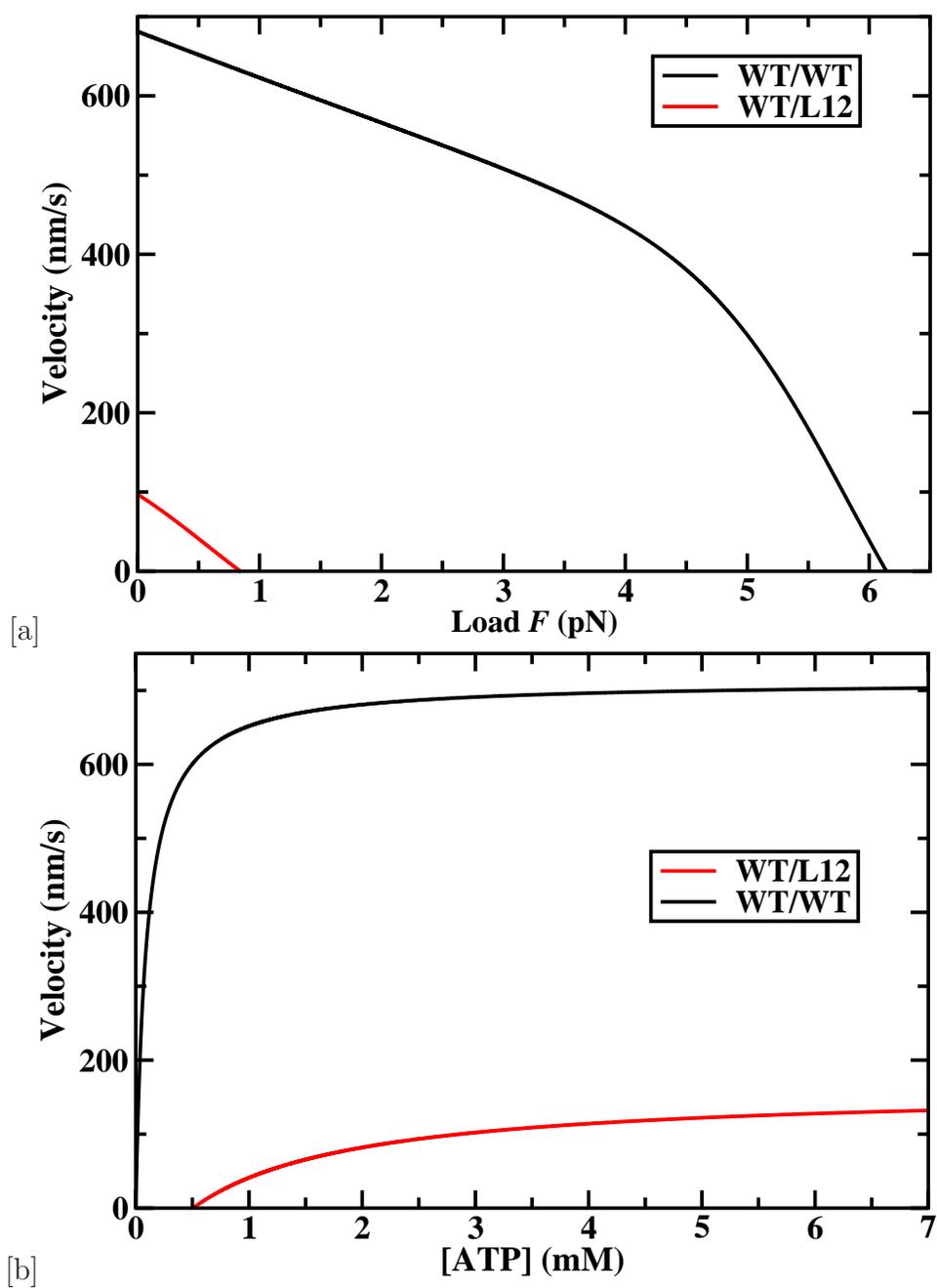

\centering
[a]\includegraphics[scale=0.5,clip=true]{Fig2a.eps}
[b]\includegraphics[scale=0.5,clip=true]{Fig2b.eps}
\caption{a) Force-velocity curves for  W/W homodimeric kinesins (black line) and for  W/M heterodimeric  kinesins (red line) at [ATP]=1 mM.  b) Velocities of  W/W homodimeric kinesins (black line) and   W/M heterodimeric  kinesins (red line) as a function of [ATP] at the constant external force $F=0.5$ pN. The velocity  of  homodimeric M/M kinesins is zero at all conditions.}
\label{fig2}
\end{figure}

In addition to the mean velocity,  an important characteristic of the motor protein dynamics is a diffusion constant or dispersion \cite{AR}. In our theoretical framework it can be calculated exactly, and the results of these computations are given in Fig. 3. The effect of external forces on dispersions for homodimeric and heterodimeric kinesins is shown Fig. 3a. External forces have qualitatively different effects on the fluctuations of wild-type and mutated motor proteins. Dispersion for homodimeric W/W kinesins is a decreasing function of the external load, while for homodimeric M/M proteins it increases. It is interesting to note that the behavior of the heterodimeric W/M is non-monotonic, depending on the strength of the external forces. At low external loads ($F \le 1$ pN) the dispersion of W/M kinesins decreases, as is for W/W kinesins, however for larger external forces it starts to increase, similarly to M/M kinesins. Dispersion as a function of [ATP] is plotted in Fig. 3b. Again, the behavior of heterodimeric W/M proteins is intermediate between wild-type and fully mutated homodimeric kinesins.

\begin{figure}[tbp]
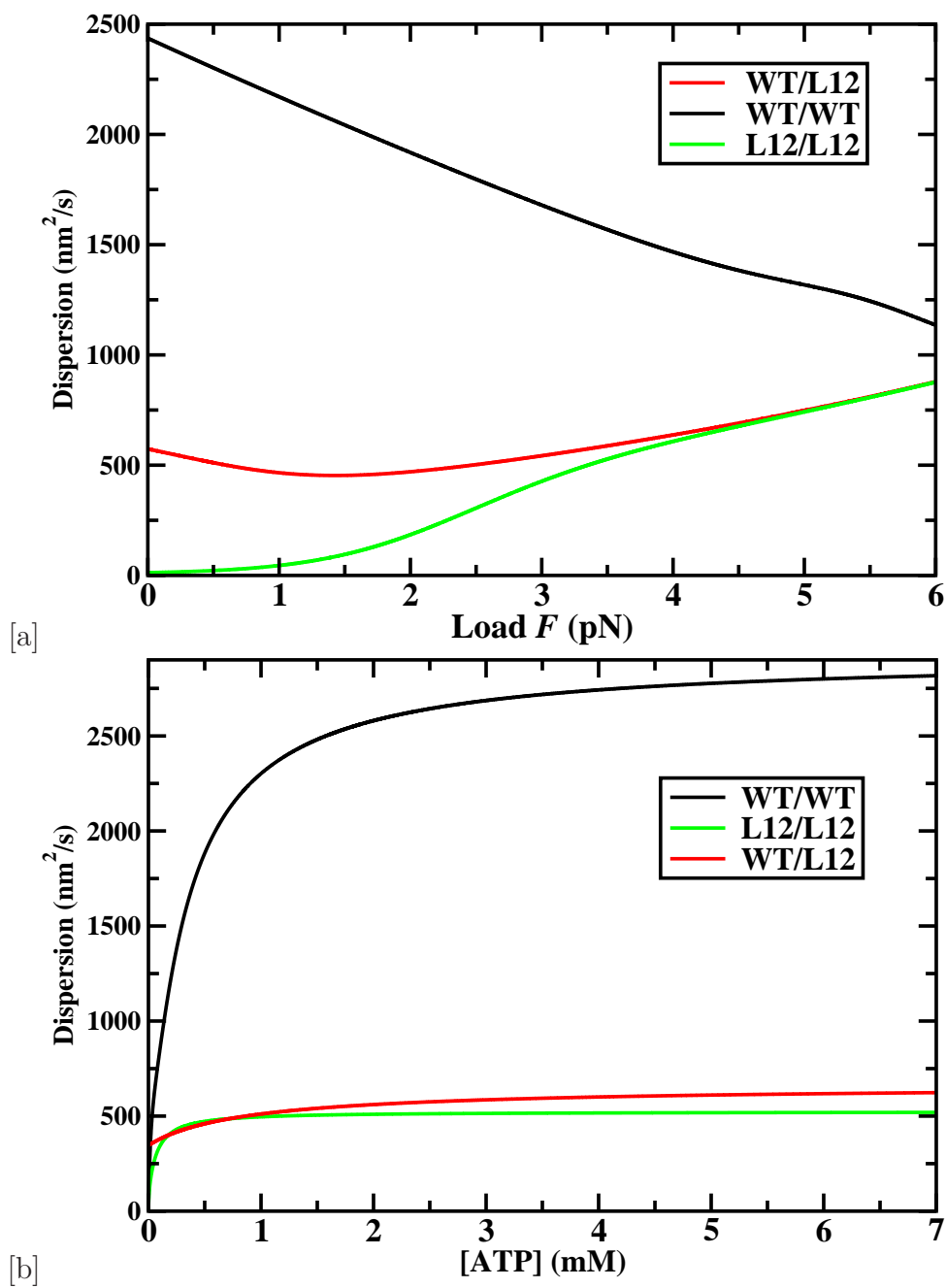

\centering
[a]\includegraphics[scale=0.5,clip=true]{Fig3a.eps}
[b]\includegraphics[scale=0.5,clip=true]{Fig3b.eps}
\caption{a) Dispersion as a function of the external force  for  W/W homodimeric kinesins (black line), for homodimeric M/M kinesins (green line),  and for  W/M heterodimeric  kinesins (red line) at [ATP]=1 mM.  b) Dispersions of  W/W homodimeric kinesins (black line), homodimeric M/M kinesins (green line), and   W/M heterodimeric  kinesins (red line) as a function of [ATP] at the constant external force $F=0.5$ pN.}
\label{fig3}
\end{figure}

Our theoretical calculations of dynamic properties and analysis of the experimental observations for homodimeric and heterodimeric kinesins suggest the following mechanistic (however strongly simplified) picture of the motor protein dynamics. Mutations change the free energy landscapes for the enzymatic activity and the mechanical processivity of molecular motors. However, motor domains in W/W, M/M and W/M proteins interact with each other differently, leading to different free energy surfaces. As a result, two heads in the heterodimeric molecule become very similar in the biochemical properties, but different from the corresponding motor domains in the homodimers. The heterodimeric motor protein still moves along the microtubules in the hand-over-hand mechanism, although with the transition rates modified by interaction between motor heads. We suggest that this theoretical picture can be tested in experiments by, for example, labeling differently both motor domains to obtain a necessary dynamic information.

\section {Conclusions}

We developed a simple theoretical description of the dynamics of motor proteins based on the discrete sequential stochastic models. This approach allows us to resolve the contradiction between experimental observations on homodimeric and heterodimeric kinesins and the widely accepted hand-over-hand stepping mechanism for two-headed molecular motors. It is argued that the interaction between motor domains can modify free energy landscapes for the motor protein motion, and the transitions rates change  depending on the nature of these domains. Explicit calculations of dynamics properties, such as velocities, dispersions, and stall forces are presented for homodimeric and heterodimeric kinesins with L12 mutations. Theoretical predictions agree well with available experimental data. Several suggestions on how to test theoretical predictions are discussed.

\ack

 The support from the Hammill Innovation Award, from the Welch Foundation (under Grant No. C-1559), and from the US National Science Foundation through the grant CHE-0237105 is gratefully acknowledged. A.B.K. thanks Prof. M.E. Fisher for valuable comments and suggestions.


\section*{References}

\begin{thebibliography}{99}      

\bibitem{lodish} Lodish H A, Berk A, Zipursky S L  and  Matsudaira P 1995 {\it Molecular Cell Biology} 3rd Ed. (New York: Scientific American Books)

\bibitem{bray} Bray D 2001 {\it Cell Movements: from molecules to motility} 2nd Ed. (New York: Garland Publishing)

\bibitem{howard} Howard J 2001 {\it Mechanics of Motor Proteins and the Cytoskeleton} (Sunderland, MA: Sinauer Associates)

\bibitem{schliwa} Schliwa M. ed. 2003 {\it Molecular Motors} (Weinheim, Germany: Wiley-VCH)

\bibitem{AR} Kolomeisky A B and Fisher M E 2007 {\it Annu. Rev. Phys. Chem.} {\bf 68} 675

\bibitem{kolomeisky05} Kolomeisky A B and Phillips H 2005 {\it J. Phys.: Condens. Matter} {\bf 17} S3887
                                                                         
\bibitem{yildiz03} Yildiz A,  Tomishige M,  Vale R D and  Selvin P R  2003  {\it Science} {\bf 302} 676

\bibitem{snyder04} Snyder G E, Sakamoto T, Hammer J A, Sellers J R and Selvin P R 2004 {\it Biophys. J.} {\bf 87} 1776

\bibitem{yildiz04} Yildiz A, Park H, Safer D, Yang Z, Chen L Q, et al.  2004  {\it J. Biol. Chem.} {\bf 279} 37223

\bibitem{toba06} Toba S, Watanabe T M, Yamaguchi-Okimoto L, Toyoshima Y Y and Higuchi H 2006 {\it Proc. Natl. Acad. Sci. USA} {\bf 103} 5741

\bibitem{reck-peterson06} Reck-Peterson S L, Yildiz A, Carter A P, Gennerich A, Zhang N and Vale R D 2006 {\it Cell} {\bf 126} 335

\bibitem{kaseda02} Kaseda K, Higuchi H and Hirose K 2002 {\it Proc. Natl. Acad. Sci. USA} {\bf 99} 16058

\bibitem{julicher97} J\"{u}licher F, Ajdari A and Prost J 1997 {\it Rev. Mod. Phys.} {\bf 69} 1269

\bibitem{bustamante01} Bustamante C, Keller D and Oster G 2001 {\it Acc. Chem. Res.} {\bf 34} 412

\bibitem{li04}  Li G and  Cui Q 2004  {\it J. Phys. Chem. B} {\bf 108} 3342 

\bibitem{hyeon07} Hyeon C and Onuchic J N 2007 {\it Proc. Natl. Acad. Sci. USA} {\bf 104} 2175

\bibitem{FK} Fisher M E  and  Kolomeisky A B  1999    {\it Proc. Natl. Acad. Sci. USA} {\bf 96} 6597 

\bibitem{kolomeisky01} Kolomeisky  A B  2001   {\it J. Chem. Phys.} {\bf 115} 7253

\bibitem{KF00a} Kolomeisky  A B  and  Fisher M E  2000   {\it Physica A} {\bf 279} 1

\bibitem{KF00b} Kolomeisky A B and  Fisher M E 2000   {\it J. Chem. Phys.} {\bf 113} 10867

\bibitem{FK2} Fisher M E  and  Kolomeisky A B  2001  {\it Proc. Natl. Acad. Sci. USA} {\bf 98} 7748 

\bibitem{KF03} Kolomeisky  A B  and  Fisher M E 2003  {\it Biophys. J.} {\bf 84} 1642

\bibitem{stukalin05} Stukalin E B, Phillips H and Kolomeisky A B 2005 {\it Phys. Rev. Lett.} {\bf 94} 238101

\bibitem{fisher05} Fisher M E and Kim Y C 2005 {\it Proc. Natl. Acad. Sci. USA} {\bf 102} 16209










\end{thebibliography}
\end{document}